\def\la{\;
\raise0.3ex\hbox{$<$\kern-0.75em\raise-1.1ex\hbox{$\sim$}}\; }
\def\ga{\;
\raise0.3ex\hbox{$>$\kern-0.75em\raise-1.1ex\hbox{$\sim$}}\; }
\documentstyle[psfig,referee]{mn}
\begin{document}
\title[Monthly Notices: GP-effect and Ly$\alpha$ forest]
{The Gunn-Peterson effect and the Lyman alpha forest}
\author[S.A. Levshakov and W.H. Kegel]
{Sergei A. Levshakov$^{1}$  and  Wilhelm H. Kegel$^{2}$\\
$^{1}$Yukawa Institute for Theoretical Physics, Kyoto University, Kyoto 606-01, Japan and \\
\hspace*{0.2cm}Department of Theoretical Astrophysics, A. F. Ioffe Physico-Technical Institute,\\
\hspace*{0.2cm}194021 St.Petersburg, Russia\\
$^{2}$Institut f\"ur Theoretische Physik der Universit\"at Frankfurt am Main, Postfach 11 19 32,\\
\hspace*{0.2cm}60054 Frankfurt/Main 11, Germany}
\maketitle
\begin{abstract}
We show that spatial correlations in
a stochastic large scale velocity field in an otherwise smooth
intergalactic medium (homogeneous comoving density)
superposed on the general Hubble flow, may cause a `line-like' structure
in QSO spectra similar to the population of unsaturated Ly$\alpha$ forest
lines which usually are attributed to individual clouds 
with 
$10^{11} \la N_{\rm HI} \la 5\times10^{13}$ cm$^{-2}$. Therefore
there is no clear observational distinction between a diffuse intergalactic medium
and discrete intergalactic clouds.
It follows that the HI-density in the diffuse intergalactic medium
might be substantially underestimated if it is determined 
from the observed intensity distribution near 
the apparent continuum in high resolution spectra of QSOs.
Our tentative estimate implies 
a diffuse neutral hydrogen opacity $\tau_{\rm GP} \sim 0.3$ at $z \sim 3$ and 
a current baryon density 
$\Omega_{\rm IGM} \simeq 0.08$, assuming a Hubble constant
${\rm H}_0 = 70$ km s$^{-1}$ Mpc$^{-1}$.
\end{abstract}

The intergalactic medium [IGM]
may be probed by observing absorption features
in the spectra of QSOs. The continuum on the blue side of the HI Ly$\alpha$
emission line is expected to be depressed as compared to that on the red side, since
Ly$\alpha$ absorption in the diffuse intergalactic medium leads to an
apparently continuous absorption due to the general cosmological
expansion (Scheuer 1965, Gunn \& Peterson 1965).
If this effect can be measured accurately,
it allows in principle  to estimate the intergalactic density of neutral
hydrogen, $n({\rm HI})$. 

In their original paper
Gunn \& Peterson [GP] estimated the depression in the spectrum of
3C9 to be of the order of 40 per cent  
and $n({\rm HI})$ to be of about $6\times10^{-11} \  {\rm cm}^{-3}$
at $z=2$ .
This low value of $n({\rm HI})$ is usually taken as an
indication for a very high degree of ionization of the IGM.

However, the measurement of the true depression 
caused by the smoothly distributed intergalactic hydrogen is hampered
by numerous absorption lines blueward of the Ly$\alpha$ 
emission line, which make it difficult to locate the continuum
level correctly and to remove the contribution to the depression 
from individual absorbers accurately.  
There have been suggested a few methods to overcome this problem and to 
measure the diffuse neutral
hydrogen absorption (for a review see e.g. Fang \& Crotts 1995).
The conclusion is that either
`{\it there is no evidence for any Gunn-Peterson effect up to the
highest redshifts observable in quasar spectra}'
( Giallongo {\it et al.} 1995), or that
$\tau_{\rm GP} = 0.115 \pm 0.025$ 
at $z \simeq 3.4$ (Fang \& Crotts 1995).
In any case the effect is small, indicating a
very low density of neutral hydrogen in the diffuse intergalactic
medium.

It is worth pointing out that in quantitative
estimates the IGM is usually considered as a thermal gas and the possible
influence of a
hydrodynamical velocity field superposed on the general Hubble flow is
ignored.
Commonly the individual Ly$\alpha$
lines being interpreted as arising in discrete intergalactic clouds
at different redshifts 
are evaluated by means of a Voigt profile 
fitting analysis (see e.g. Lu {\it et al.} 1996).
In this procedure
there are three free parameters, the column density $N_{\rm HI}$,
the Doppler parameter $b$,
and the redshift $z_a$, 
to be determined from the observational data. 
Implicitly it is also assumed
that any hydrodynamical velocities within the individual clouds may be
accounted for
in the {\it microturbulent} approximation in which any bulk motions are treated
as
completely uncorrelated (i.e. the velocity correlation length $l_0 \equiv 0$). 

Thus, in the standard analysis it appears observationally possible to
{\it clearly distinguish}  between the Ly$\alpha$  absorption in the diffuse
intergalactic medium leading to a true
GP-effect and that in the individual clouds leading to the Ly$\alpha$  forest lines.
As will be shown below, this
conclusion is misleading and is a consequence of the simplifying
assumptions underlying the common approach. As a result the diffuse neutral hydrogen
opacity mentioned above may have been considerably underestimated.

With respect to the saturated
Ly$\alpha$  forest lines, corresponding to clouds with 
$10^{14} \la N_{\rm HI}$ $\la 10^{16}$
cm$^{-2}$ it has been pointed out
recently (Levshakov \& Kegel 1996), that the
quantitative interpretation may be changed substantially when one
accounts for
a stochastic velocity  field with finite correlation length within the
individual clouds. In particular it has been shown by means of
Monte-Carlo [MC]
simulations (Levshakov, Kegel \& Mazets 1997)
that such a velocity field can lead to a rather structured line
profile with several subcomponents, which in the traditional analysis would
be attributed to individual clouds while in the underlying model a
homogeneous density was assumed.

This latter result led us to investigate in which way the GP-effect
is modified when a stochastic velocity field
with a finite correlation length is superposed on the general Hubble
flow with a homogeneous comoving density. 
In fact our results show (Fig.~1) that such fields can give rise
to a `line-like' structure
of absorption features similar to the observed unsaturated
Ly$\alpha$  forest lines 
($10^{11} \la N_{\rm HI} \la 5\times10^{13}$ cm$^{-2}$).

\begin{figure}
\vspace{-3.5cm}
\centerline{\vbox{
\psfig{figure=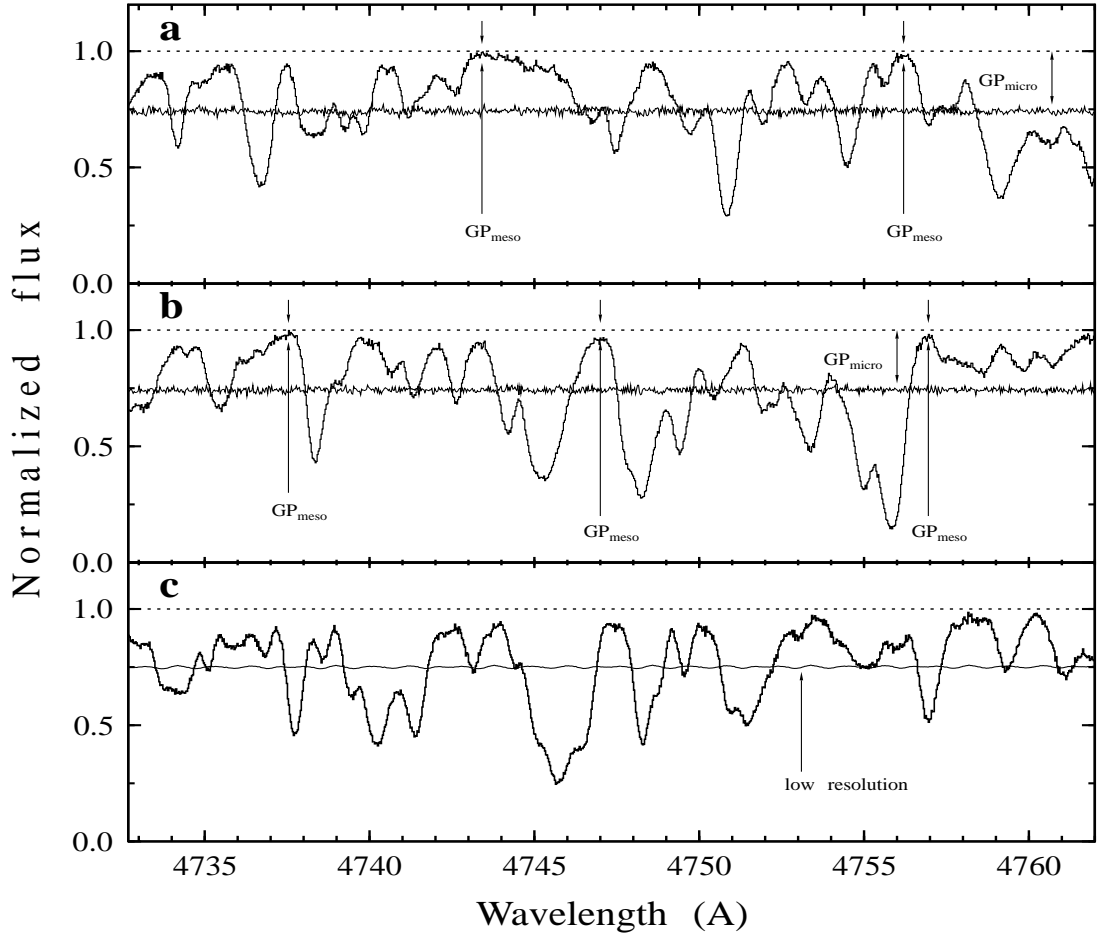,height=18.0cm,width=17.0cm}
}}
\vspace{-1.0cm}
\caption[]{MC simulations of QSO absorption spectra
for three random realizations of the stochastic velocity field along a given
line-of-sight.
Panels {\bf a}, {\bf b}, and {\bf c} show a portion of the spectrum
blueward of Ly$\alpha$ emission for $z_e = 3$ computed with the parameters
$q_0 = 0.5$, ${\rm H}_0 = 70$ km s$^{-1}$ Mpc$^{-1}$,
and
$n({\rm HI}) = 4\times10^{-11}\ {\rm cm}^{-3}$.
In addition to the {\it mesoturbulent}
spectra, results obtained
under the condition that the bulk motions have
zero correlation coefficient and $\sigma_{t}/v_{th} > 1$
(panel {\bf a}), or
have a finite correlation coefficient but
$\sigma_{t}/v_{th} \ll 1$ (panel {\bf b}) are shown in order to
demonstrate the {\it microturbulent} solution and
the case of purely thermal broadening, respectively.
Panel {\bf c} shows both a high-resolution and a low-resolution spectrum
with FWHM = 0.12 \AA\   and  80 \AA, respectively.
The number density of the absorption features in this portion
of the simulated QSO spectrum is $N(z) \simeq 500$ which
is in a good agreement with observations
(cf.  Kirkman \& Tytler 1997).
}
\end{figure}

\begin{figure}
\vspace{-2cm}
\centerline{\vbox{
\psfig{figure=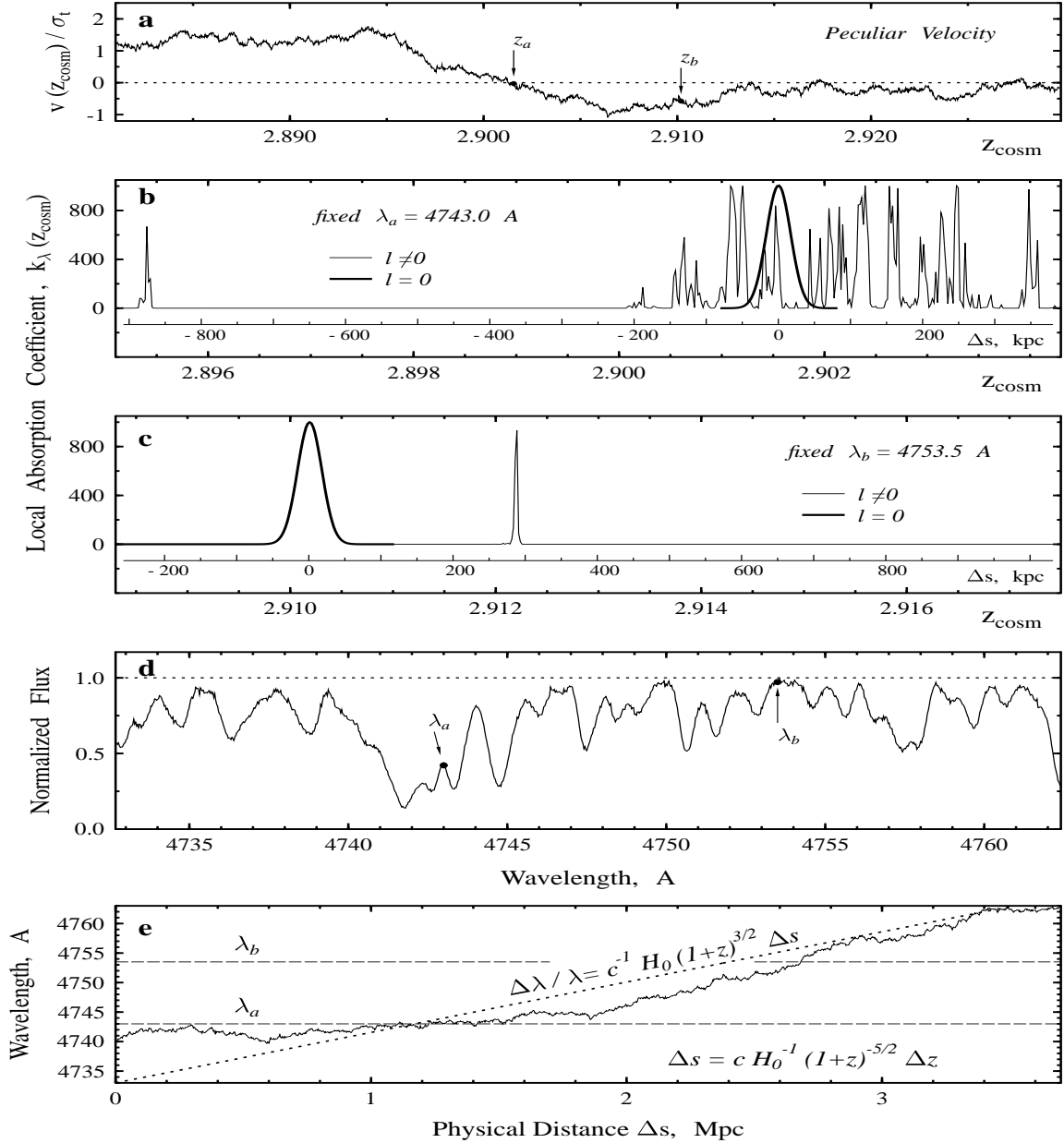,height=18.0cm,width=16.0cm}
}}
\vspace{1cm}
\caption[]{An example of MC simulations.
({\bf a}) -- One realization of the stochastic velocity field
(in units of the rms velocity $\sigma_t$)
versus the cosmological redshift $z_{cosm}$.
({\bf b}) and ({\bf c}) --
The corresponding local absorption coefficients for fixed
$\lambda_a = 4743$ \AA\  and $\lambda_b = 4753.5$ \AA,\  respectively 
(thin curves, correlation length
$l \neq 0$), and for comparison $k_{\lambda}$
for a purely thermal gas (thick curves, $l = 0$ ).
({\bf d}) -- A portion of the resulting absorption spectrum. The simulated
intensities at the fixed wavelengths
$\lambda_a$ and $\lambda_b$ are marked by dots and the corresponding 
cosmological redshifts
$z_a$ and $z_b$ are shown on panel ({\bf a}). Panel ({\bf e}) illustrates the effect
of the contribution from different volume elements to the absorption at a given value
of $\lambda$ when the peculiar velocity field (panel {\bf a}) is superposed
on the general Hubble flow.
}
\end{figure}

\begin{figure}
\centerline{\vbox{
\psfig{figure=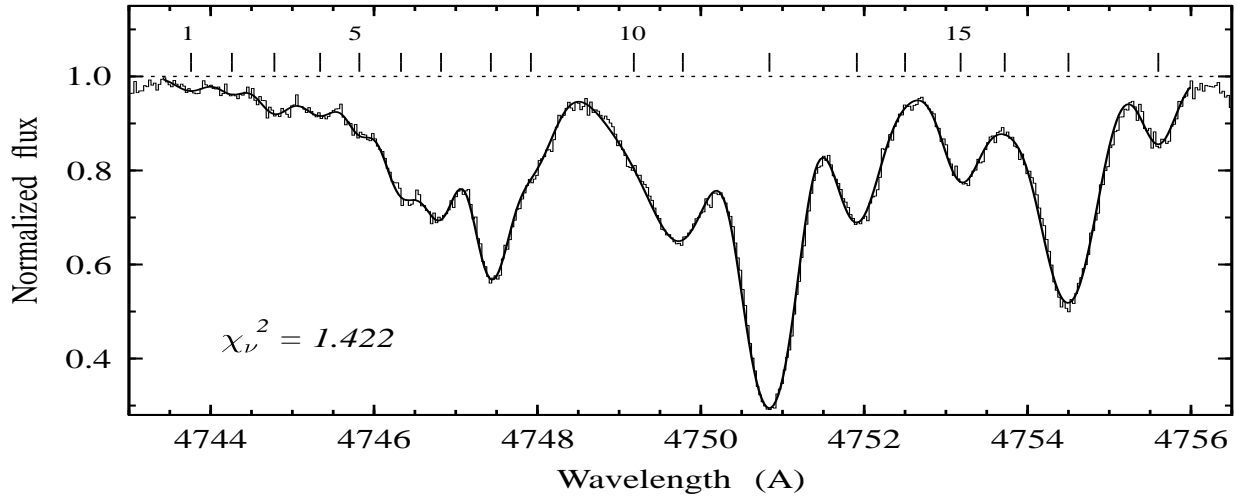,height=8.0cm,width=12.5cm}
}}
\vspace{-1cm}
\caption[]{A section of the simulated  Ly$\alpha$ forest
spectrum shown in Fig.~1a (thin curve) fitted by the 
superposition of 18 Voigt profiles (thick curve). 
The average deviation of the two curves 
($\chi_{\nu}^2$ per degree of freedom) is indicated.  }
\end{figure}

To describe our {\it mesoturbulent} model $(l_0 > 0)$
the following set of the parameters  has to be specified :
the neutral hydrogen number density $n({\rm HI})$ at cosmic time $t$, 
the kinetic temperature $T_{kin}$,
the ratio of the rms value of the line-of-sight stochastic
velocity dispersion $\sigma_{t}$
to the thermal velocity $v_{th}$, the correlation length $l_0$, and
the parameter $q_0$.
We take the Hubble constant in units of 
${\rm H}_0$ = 70 km s$^{-1}$ Mpc$^{-1}$
and the cosmological constant
$\Lambda = 0$.
The stochastic velocity field is assumed to be governed by 
a stationary Markov process (see Levshakov \& Kegel 1994, 1997 for details).

The value of $\sigma_{t}$ may be estimated from the observed
distribution function of
peculiar velocities of galaxies. Recent observations indicate, for example,
large-scale motions
with $\sigma_{gal} \sim 700$ km s$^{-1}$ on scales up to 70 Mpc
(Raychaudhury \& Saslaw 1996).
However, if the peculiar bulk motions are produced by gravity
they should be smaller at high redshifts.
An order of magnitude of $\sigma_{t}$ at $z = 3$ may be estimated from
the linear theory of the gravitational instability which leads to
$\sigma_{t} \propto (1+z)^{-1/2}$ when the density parameter
$\Omega \equiv 2q_0$ is equal to 1 (see e.g. Zel'dovich \& Sunyaev 1980).
For our MC simulations we therefore adopt
$\sigma_{t} = 300\ {\rm km\ s}^{-1}$, as well as  $q_0 = 0.5$
and $z_e = 3$. The choice of the kinetic temperature of
$T_{kin} = 10^4$ K is based on the arguments given by Levshakov \& Kegel (1996).
For the correlation length $l_0$ we chose a
value of 24 Mpc. Assuming the velocity fluctuations to be
`frozen' into the IGM and just expanding along with the general expansion,
we then have $l_z = l_0/(1+z) = 6$ Mpc.

An order of magnitude of $n({\rm HI})$ may be estimated from the average
value of $D_A$ -- the continuum depression between HI Ly$\alpha$ and Ly$\beta$ 
emission lines in low dispersion spectra
with FWHM$ \sim 80-100$ \AA\
[$D_A \equiv \langle 1 - F_{obs}/F_{int} \rangle$, where $ F_{obs}$
and $F_{int}$
are the observed and intrinsic fluxes per unit wavelength in the QSO rest
frame]. 
At $z = 3$ one finds $D_A \simeq 0.25$ according to the data compiled by
Kennefick {\it et al.} (1995).

We will consider in some detail a narrow spectral range of 
$\Delta \lambda_{obs} = 30$ \AA\  between the Ly$\alpha$ and $\beta$
emission lines. For this range we 
assume that the contribution to $D_A = 0.25$ comes in general
from the cumulative effect of weak absorption features, and we will show
that in this case the Doppler effect caused by the peculiar velocities
superposed on the Hubble flow leads to the observed picture --
instead of a characteristic smooth GP-trough one observes a `line-like'
structure. We emphasize that this does not rule out the possibility of
a clumpy IGM, but  our results indicate that the interpretation of the current high
resolution observations of QSOs is not unique and therefore the classical analysis may yield
misleading parameters of the IGM.
 
Having specified the free parameters we can simulate a QSO absorption
spectrum for a random realization of the velocity field.
All our synthetic spectra were calculated with a signal-to-noise ratio of 100, 
an instrumental resolution of 8 km s$^{-1}$,
and a pixel size of 2 km~s$^{-1}$ to match the data obtainable with
the HIRES spectrograph of the Keck I telescope.

We calculate $\tau_\lambda$,
the total optical depth at wavelength $\lambda$, 
following GP, the essential difference being that in our model the local
absorption coefficient $k_{\lambda}(z)$ is a stochastic variable (see Fig.~2b,c)
which we
estimate by means of our MC technique
(Levshakov, Kegel \& Mazets 1997).

Fig.1 shows three mesoturbulent spectra
calculated for three different realizations of the velocity field.
For comparison and in order
to validate our procedure, we calculated in addition
$\tau_\lambda$ for zero correlation coefficient $f = 0$
and $\sigma_{t}/v_{th} > 1$,
and for a finite correlation coefficient but $\sigma_{t}/v_{th} \ll 1$. 
For both cases one
expects to obtain a classical GP-trough. The first case corresponds to the 
microturbulent limit, while in the second case thermal broadening dominates.
The results of these tests are shown in Fig.~1a and 1b, respectively, 
and marked by `${\rm GP}_{\rm micro}$'.

With $n({\rm HI}) = 4\times10^{-11}\ {\rm cm}^{-3}$ 
at $z = 3$ our microturbulent results are found to fit
the continuum depression $D_A \sim 0.25$. 
The corresponding GP-optical depth $\tau_{\rm GP}$ is
about 0.3. -- Fig.~1c shows a high resolution mesoturbulent
spectrum and for comparison the same
spectrum degraded to a resolution of 80 \AA. The latter resembles the
microturbulent spectra in panels {\bf a} and {\bf b}. --
Commonly one attempts to estimate $n({\rm HI})$ in the diffuse IGM,
i.e. inbetween Ly$\alpha$
clouds, by measuring the intensity distribution 
near the apparent continuum level 
in high resolution spectra of QSOs
(see e.g. Fang \& Crotts 1995).
The  corresponding regions in our mesoturbulent spectra are labeled 
by `${\rm GP}_{\rm meso}$'  in Fig.~1.
It is clearly seen that instead of $\tau_{\rm GP} \simeq 0.3$ one may
obtain only an upper limit or a very small value 
for $\tau_{\rm GP}$ in this case.	

In order to illustrate these calculations,
Fig.~2a shows one random realization
of the stochastic velocity field along a given line of sight. --
The peculiar velocities have the effect that {\it different} volume elements
may contribute to the absorption at the {\it same} value of $\lambda$.
This is shown in Fig.~2b,c  by the thin curves. For example, the
contribution to $k_{\lambda_a}$ comes from individual volume elements 
spread over 1 Mpc.
For comparison also the absorption by a purely
thermal gas (no peculiar correlated velocity field) is shown by the thick
curves. Finally Fig.~2d shows the resulting absorption spectrum and 
Fig.~2e the effect of compensating the general Hubble expansion by
hydrodynamic perturbations. 

In order to stress the difference in the interpretation of weak
Ly$\alpha$ forest lines we fitted a part of the simulated spectrum
($4743 < \lambda < 4756$ \AA)  
shown in Fig.~1a in the usual way by considering 18 individual Voigt
profiles. The result is shown in Fig.~3 and the corresponding
parameters are listed in Table~1. As can be seen from the figure
the fit is satisfactory, the derived `physical parameters'
are in the ranges $6.6\times10^{11} \leq N({\rm HI}) \leq 4.0\times10^{13}$
cm$^{-2}$, and $12.95 \leq b \leq 37.12$ km s$^{-1}$ (note the thermal
width in our model corresponds to 12.89 km s$^{-1}$). So, one can obtain
a typical spread of `physical parameters'
for the Ly$\alpha$ forest lines (cf. e.g. Kirkman \& Tytler 1997),
while a homogeneous model was assumed.

\setcounter{table}{0}
\begin{table}
\centering
\caption{`Ly$\alpha$ cloud parameters' derived from the spectrum shown in 
Fig.~3 by fitting Voigt profiles }
\label{tb1}
\begin{tabular}{lccc}
\hline
 No. &  $\lambda_{obs}$, \AA & $N({\rm HI}),$ cm$^{-2}$ & $b$, km s$^{-1}$\\ \hline
1  & 4743.76 & 6.93(11) & 16.10 \\
2  & 4744.26 & 6.65(11) & 12.95 \\
3  & 4744.78 & 1.84(12) & 16.56 \\
4  & 4745.34 & 2.16(12) & 18.50 \\
5  & 4745.82 & 2.03(12) & 13.13 \\
6  & 4746.33 & 7.21(12) & 18.99 \\
7  & 4746.82 & 7.89(12) & 17.32 \\
8  & 4747.43 & 1.28(13) & 17.88 \\
9  & 4747.92 & 6.60(12) & 22.78 \\
10 & 4749.18 & 6.78(12) & 37.12 \\
11 & 4749.78 & 1.53(13) & 30.43 \\
12 & 4750.84 & 4.02(13) & 24.54 \\
13 & 4751.91 & 1.15(13) & 23.01 \\
14 & 4752.50 & 1.10(12) & 21.12 \\
15 & 4753.18 & 6.33(12) & 19.30 \\
16 & 4753.72 & 3.00(12) & 23.51 \\
17 & 4754.50 & 2.32(13) & 26.43 \\
18 & 4755.60 & 3.63(12) & 17.05 \\ \hline
\end{tabular}
\end{table}
 
Thus, our calculations offer a new interpretation of the observed
Ly$\alpha$ forest
lines and explain why high resolution spectra show
only a very small  GP-effect.
Our results show that at least some 
of the Ly$\alpha$ forest lines may be
caused by absorption in the diffuse medium between the intervening
clouds giving rise to saturated lines. This implies that in the diffuse IGM
$n({\rm HI})$
may be substantially higher than estimated from the upper limits for
$\tau_{\rm GP}$ in previous studies.
For example Giallongo {\it et al.} (1995) give a value
$\tau_{\rm GP} = 0.01 \pm 0.03$ for PKS 2126--158 ($z_e = 3.3$) which implies
$n({\rm HI}) \la 2\times10^{-12}$ cm$^{-3}$. This is an order of magnitude
below the value we assumed in our model. -- If one wants to derive from
$n({\rm HI})$ the total baryon density,
one has to know the degree of ionization
of the IGM. With the assumptions that
(i) the ionization of the IGM is in equilibrium,
(ii) the ionizing continuum can be described by 
a power law with spectral index = $-1.5$ and a flux
at the hydrogen Lyman limit 
$ \simeq 3\times10^{-21}$ ergs s$^{-1}$ cm$^{-2}$ sr$^{-1}$ Hz$^{-1}$ 
(see e.g. Bechtold 1994),
(iii) $n({\rm HI}) = 4\times10^{-11}$ cm$^{-3}$,
(iv) $T_{kin} = 10^4$ K, and
(v) all the IGM density is in the form of hydrogen,
an order of magnitude estimate leads
to a current baryon density of $\Omega_{\rm IGM} \simeq 0.08$, i.e.
it is about 8 per cent of the mass needed to close the
Universe.

Taking this at face value, we conclude that
$\Omega_{\rm IGM} \simeq 0.08$ is consistent with the most conservative
upper limit on baryonic mass $\Omega_b \la 0.06$ from Big Bang
Nucleosynthesis (e.g. Fields \& Olive 1996).
It follows that the density of the IGM may be considerably larger than the
observed density of luminous matter $0.003 \la \Omega_{lum} \la 0.007$
(e.g. Jedamzik {\it et al.} 1995), and the baryon density in the
Ly$\alpha$ clouds, $\Omega_{\rm Ly} \simeq 0.002 - 0.003$
(Lanzetta {\it et al.} 1991). 

While our model is highly idealized (smooth density),
the results clearly show that the formation of the GP-depression
is intimately related
to the formation of narrow absorption lines.
Thus, there is no simple way to
distinguish observationally between the diffuse IGM and intervening
clouds as
was suggested by the classical interpretation of the
GP-effect and the Ly$\alpha$ forest lines.

\vspace{5mm}

{\bf Acknowledgements}

\medskip\noindent
We thank K. Tomita and F. Takahara for discussions 
and an anonymous referee for constructive criticism.
This work was supported in part by the RFFR grant
No. 96-02-16905-a and by the Deutsche Forschungsgemeinschaft.
Numerical computations
in this work was supported by the Yukawa Institute for Theoretical Physics.

\end{document}